\documentclass[graybox]{svmult}

\usepackage{type1cm}        
\usepackage{makeidx}         
\usepackage{graphicx}        
\usepackage{multicol}        
\usepackage[bottom]{footmisc}

\usepackage{newtxtext}       %
\usepackage[varvw]{newtxmath}       

\makeindex             

\renewcommand{\sl}{\mathfrak{sl}}
\newcommand{\g}{\mathfrak{g}}
\newcommand{\C}{\mathbb C}
\newcommand{\Z}{\mathbb Z}
\newcommand{\bx}{\mathbf x}
\newcommand{\bs}{\mathbf s}
\newcommand{\al}{\alpha}
\newcommand{\cD}{\mathcal D}
\newcommand{\cH}{\mathcal H}
\newcommand{\half}{\frac12}

\begin{document}

\title*{Duality and Macdonald difference operators}
\author{Philippe Di Francesco and Rinat Kedem}
\institute{Philippe Di Francesco \at University of Illinois, Mathematics MC-382 Urbana IL 61821 USA
and IPhT CEA, Universit\'e Paris-Saclay 91191 Gif-sur-Yvette France, \email{philippe@illinois.edu}
\and Rinat Kedem \at University of Illinois, Mathematics MC-382 Urbana IL 61821 USA, \email{rinat@illinois.edu}}
%
%
\maketitle

\abstract{This note summarizes certain properties common to Macdonald, Koornwinder and Arthamonov-Shakirov $q$-difference operators, relating to the duality or bi-spectrality properties of their eigenfunctions. This results in Pieri operators which, in the $q$-Whittaker limit, are relativistic difference Toda type Hamiltonians which have a related quantum cluster algebra structure known as the quantum Q-system. The genus-2 result explained here is new.}

\section{Introduction}
\label{sec:1}
This contribution puts together certain results associated with the duality or bi-spectrality property which is common to the (genus 1) spherical double affine Hecke algebras (sDAHAs) of classical types and the recently introduced genus-2 DAHA \cite{ArtShak}. The functional representation of the spherical DAHA contains a set of distinguished generators which are the  (generalized) Macdonald/Koornwinder $q$-difference operators. In genus 1, Macdonald operators are naturally related to the affine root system of $A_{N}^{(1)}$, while Koornwinder's operators are related to $BC_N$ type affine
root systems. In genus-2, only the rank-1 DAHA is defined \cite{ArtShak}, and we propose a candidate which takes the place of the affine root system in defining the difference operators and duality. 
The duality relates the Macdonald or Koornwinder $q$-difference operators with the Pieri rule operators, which are difference operators in the ``weight variables" $\lambda$. 

In the $q$-Whittaker limit when $t\to\infty$, we have shown \cite{DFK21} that in the genus-1 case, the Dehn twists of the distinguished generators of the sDAHAs are $A$-type cluster variables in the quantum Q-system cluster algebras associated to each root system. The Pieri operators are interpreted as  Toda-type Hamiltonians, conserved quantities of the discrete evolution given by appropriate mutation sequences.
We show below that the genus-2 theory degenerates in this limit into a product of three genus 1 $A_1^{(1)}$-type quantum Toda theories.

We also discuss the role of the Universal solutions (c.f. basic Harish-Chandra series \cite{Stokman}), in terms of which the duality takes a particularly simple form, both in the sDAHA setting and in the $q$-Whittaker limit. 
\bigskip

\noindent{\bf Acknowledgments.}
We thank the organizers of the program ``Geometric and Representation-Theoretic Aspects of Quantum Integrability," Simons Center for Geometry and Physics in Stony Brook, and M. Shapiro for discussions about genus-2 DAHA.
RK thanks the Institut de Physique Th\'eorique, Saclay for hospitality. 
Research supported by NSF  DMS  grant 18-02044 and the Morris and Gertrude Fine endowment.

\section{Duality for type $A$ Macdonald theory}
\label{sec:2}

Macdonald theory is associated with $A_{N-1}^{(1)}$ or affine $\sl_N$. 
%
The finite-dimensional algebra $\sl_N$ has simple roots $\Pi=\{\alpha_i=\epsilon_i-\epsilon_{i+1}, i=1,...,N-1\}$ and positive roots $R_+ = \{\epsilon_i-\epsilon_j : 1\leq i < j \leq N\}=-R_-$. The fundamental weights are $\omega_i=\epsilon_1 + \cdots + \epsilon_i$.
Let $Q_+$ denote the non-negative integer cone generated by $\Pi$.
The Weyl group $W=S_N$ acts by permutation of the  basis vectors $\epsilon_i$ of $\C^N$.
With $q$ a generic complex number with $|q|<1$,
the quantum torus or $q$-Weyl algebra $\mathbb T_X$ is generated by $N$ commuting variables $x_1,...,x_N$ and and $N$ commuting momenta $\Gamma_1,...,\Gamma_N$, subject to the relations
\begin{equation}\label{qgamax} 
\Gamma_i\, x_j =q^{\delta_{i,j}}\, x_j \, \Gamma_i.
\end{equation}
In the functional representation, $x_i$ acts on functions of $\bx=(x_1,...,x_N)$ by  multiplication and $\Gamma_i$ by shift of $x_i$:
$\Gamma_i f(\bx)=f(x_1,...,q x_i, ...,x_N)$, i.e. $\Gamma_i=q^{x_i\partial_{x_i}}$.

\subsection{Macdonald operators and eigenfunctions}

Let $\vartheta(x)$ the Jacobi theta function
\begin{eqnarray*}
\vartheta(x)=
\frac{1}{2i} q^{1/8} x^{-\half} (q;q)_\infty (x;q)_\infty (qx^{-1};q)_\infty,
\end{eqnarray*}
so that
$$
 \frac{\vartheta(x;q)}{\vartheta(tx;q)} = t^{\half}\prod_{n\geq 0} \frac{1-q^n x}{1-tq^n x}\prod_{n>0} \frac{1-q^n x^{-1}}{1-t^{-1}q^n x^{-1}}.
$$
Define the  function
$$
\Theta(\bx) = t^{-{N \choose 2}} \prod_{\alpha\in R_+}\frac{\vartheta(\bx^\alpha)}{\vartheta(t \bx^\alpha)} = t^{-{N \choose 2}}\Delta_+(\bx) \Delta(\bx),
$$
where $\bx^{\epsilon_i}=x_i$, and
$$
\Delta_+ (\bx)= \prod_{n\geq 0}\prod_{\alpha\in R_+}\frac{1-q^n \bx^\alpha}{1-t q^n \bx^\alpha},\qquad
\Delta(\bx) = \prod_{n> 0}\prod_{\alpha\in R_-}\frac{1-q^n \bx^\alpha}{1-t^{-1} q^n \bx^\alpha}.
$$
The modular property of the theta function $q x \vartheta(qx)=\vartheta(x)$ implies that
$$
\Theta(\bx) \Gamma_{\omega_a} \Theta(\bx)^{-1} = t^{-a(N-a)} \Gamma_{\omega_a},
$$
where $\Gamma_{\omega_a}=\Gamma_1\cdots \Gamma_a$.
Therefore,
$$
\Delta_+(\bx)^{-1}\Gamma_{\omega_a} \Delta_+(\bx) = t^{a(n-a)} \Delta(\bx)\Gamma_{\omega_a} \Delta(\bx)^{-1}=\prod_{i=1}^a \prod_{j=r+1}^N \frac{t x_i-x_j}{x_i-x_j} \Gamma_{\omega_a}= A_a(\bx) \Gamma_{\omega_a}
$$
\begin{remark} The product $\Delta_+(\bx)$ is the one introduced by Macdonald, see e.g. \cite{macdoroot}, and the discussion above shows that it can be interchanged with $\Delta(\bx)$ up to a power of $t$. The choice of $\Delta(\bx)$ is adapted to taking the $t\to\infty$ $q$-Whittaker limit.
\end{remark}

The Macdonald commuting difference operators are \cite{macdo}:
\begin{equation}\label{macdop}
\cD_a(\bx)= {\rm Sym}( A_a \Gamma_{\omega_a})=\sum_{I\subset [1,N]\atop |I|=a} 
\prod_{i\in I,j\not\in I} \frac{t x_i -x_j}{x_i-x_j} \prod_{i\in I}\Gamma_i
\end{equation}
where $\rm Sym$ stands for symmetrization over the Weyl group $S_N$ acting on $x_i$.
%
The Macdonald polynomials $P_\lambda(\bx)$ are defined as their common monic eigenfunctions:
\begin{equation}\label{eigen}
\cD_a(\bx)\, P_\lambda(\bx)= t^{-{a\choose 2}} e_a(\bs) \, P_\lambda(\bx),\qquad P_\lambda(\bx)=\bx^\lambda (1+O(\{\bx^{-\al}\}_{\al\in Q_+})), 
\end{equation}
indexed by integer partitions $\lambda=(\lambda_1\geq \lambda_2\geq \cdots \geq \lambda_N\geq 0)$.
%
Here, $\bs=q^\lambda\,t^\rho=(s_1,s_2,...,s_N)$,  $s_i=q^{\lambda_i} t^{N-i}$, with $\rho$ the Weyl vector, while $e_a(\bs)$ is the $a$th elementary symmetric function in $s_1,...,s_N$.

\subsection{Duality and Pieri rules}
The Macdonald polynomials have a remarkable duality or bi-spectrality property under the exchange of the variables $\bx$ and $\bs$. When the variables are specialized as 
$\bx=q^{\mu}\, t^\rho$  and $\bs=q^\lambda\, t^\rho$ where $\mu, \lambda$ are integer partitions, the duality property is due to Macdonald \cite{macdo}:
\begin{equation}\label{duapol}
\frac{P_\lambda(t^\rho q^\mu)}{P_\lambda(t^\rho )}=
\frac{P_\mu(t^\rho q^\lambda)}{P_\mu(t^\rho )} .
\end{equation}
Moreover, the denominators $P_\lambda(t^{\rho})$ can be expressed in terms of $\Delta(\bx)$ (c.f. \cite{macdo} (6.11)):
\begin{equation}\label{norm}
P_\lambda(t^\rho) = t^{(\rho,\lambda)} \frac{\Delta(t^\rho)}{\Delta(t^\rho q^\lambda)}
\end{equation}

The duality property can be generalized to the Universal solution \cite{ShiraishiNoumi} of the Macdonald eigenvalue equations \eqref{eigen}, also called
asymptotically-free Harish-Chandra series \cite{Stokman}. 
The polynomials $P_\lambda(\bx)$ are replaced by formal series solutions $P(\bx;\bs)=q^{(\lambda,\mu)}\sum_{\al\in Q_+} c_\al(\bs) \bx^{-\al}$ where
$\lambda$ and $\mu$ are now generic complex $N$-tuples, with the normalization $c_0(\bs)=1$. 
The series is convergent in the domain $|\bx^{-\al_i}=x_{i+1}/x_i|<1$
and truncates to a finite sum when $\lambda$ is specialized to an integer partition, recovering 
\begin{equation}\label{recover}
P(\bx;\bs=q^\lambda t^\rho)\Big\vert_{  \hbox{$\lambda$ integer partition}}=t^{-(\rho,\lambda)}\,P_\lambda(\bx) ,
\end{equation} 
by using the relation $\bx^\lambda=q^{(\lambda,\mu)}\, t^{(\rho,\lambda)}$. The duality relation \eqref{duapol}  can be generalized to the universal solution with generic $\bx, \bs$
\cite{DFK21}:
\begin{equation}\label{duauniv}
 \Delta(\bs)\, P(\bx;\bs)=\Delta(\bx)\, P(\bs;\bx) .
 \end{equation}

The Pieri rules for Macdonald polynomials follow from the duality property. These express the result of multiplication of $P_\lambda(\bx)$ by the elementary symmetric functions $e_a(\bx)$:
\begin{equation}
\label{pieri}
e_a(\bx)\, P_\lambda(\bx)=\cH_a(\bs) \, P_\lambda(\bx)
\end{equation}
Here $\cH_a(\bs)$ is a difference operator acting on functions of $\bs=(s_1,...,s_N)$:
\begin{equation}\label{pieriop} 
\cH_a(\bs) = t^{a \choose 2} {\rm Ad}_{\Delta(\bs)^{-1}\,t^{(\rho,\lambda)}} \cD_a(\bs).
\end{equation}
This follows from the eigenvalue equations for the universal solutions and \eqref{duauniv}:
$$ 
\Delta(\bx)^{-1}\cD_a(\bx)\, \Delta(\bs) P(\bx;\bs)=\Delta(\bx)^{-1}\,\cD_a(\bx)\,\Delta(\bx)\, P(\bs;\bx)=t^{-{a\choose 2}}e_a(\bs)\, P(\bs;\bx),
$$
after interchanging the names of variables $\bx\leftrightarrow \bs$, then specializing to integer partitions $\lambda$ and using \eqref{recover}.
Explicitly, Equation \eqref{pieriop} is \cite{macdo}
$$
\cH_a(\bs)=\sum_{I\subset [1,N]\atop |I|=a}
\prod_{i\in I,\ j\not \in I\atop j<i} \frac{t^{i-j-1}\Lambda_{j}-\Lambda_i}{t^{i-j}\Lambda_{j}-\Lambda_i}\ \frac{t^{i-j+1} \Lambda_j-q \Lambda_{i}}{t^{i-j}\Lambda_j-q\Lambda_{i}}\, 
\prod_{i\in I}T_i,
$$
in terms of the quantum torus $\mathbb T_\Lambda$ variables $\Lambda=q^\lambda$ and $T_i=q^{\lambda_i\partial_{\lambda_i}}$, which generate a second $q$-Weyl algebra with 
$T_i \Lambda_j = q^{\delta_{ij}} \Lambda_j T_i$.

\subsection{Dehn twist acting on sDAHA}
The operators \eqref{macdop} are generators in the functional representation of the sDAHA, i.e. 
the elementary symmetric functions of the generators $Y_i$ of DAHA. This has a natural
torus modular group  $\mathsf{SL}_2(\Z)$-action generated by the two (a and b-cycle) Dehn twists.
The Dehn twist $\tau_+$, corresponding to $\begin{pmatrix} 1 & 1 \\ 0 & 1\end{pmatrix}$ in the defining representation, acts on the functional representation 
via the adjoint action of Cherednik's Gaussian $\gamma(\bx)$ \cite{cheredbook}:
\begin{equation}\label{gaussian}
\gamma(\bx) = \exp\left\{ \frac{\sum_{i=1}^N (\log x_i)^2}{2 \log q}\right\} .
\end{equation}
The action on Macdonald operators allows to define a discrete time evolution for $n\in \Z$:
\begin{equation}\label{translatedmacdonald}
\cD_{a,n}(\bx)=q^{-na/2}\, {\rm Ad}_{\gamma^{-n}} (\cD_a(\bx))=\sum_{I\subset [1,N]\atop |I|=a} 
\prod_{i\in I,j\not\in I} \frac{t x_i -x_j}{x_i-x_j} \prod_{i\in I} x_i^n \, \prod_{i\in I}\Gamma_i
\end{equation}
In the limit $N\to\infty$ these can be identified as elements in the quantum toroidal algebra of ${\mathfrak gl}_1$ \cite{DFKqt}.

\section{Duality for Koornwinder theory}

The root system underlying Koornwinder theory is the unreduced $BC_N$ system:
$$
R_+ = \left\{\epsilon_i \pm \epsilon_j (1\leq i\leq j\leq N); \epsilon_i, 2\epsilon_i (1\leq i\leq N)\right\}.
$$
We denote by $Q_+$ the non-negative integer cone generated by the finite positive roots of type $C_N$.
The Weyl group $W\simeq \Z_2^{\times N} \rtimes S_N$ acts on the variables $\bx=(x_1,...,x_N)$  by permutations and inversion of variables.
The Weyl-invariant elementary symmetric functions are denoted by $\hat e_a(\bx)=e_a(x_1,x_1^{-1},...,x_N,x_N^{-1})$.

\subsection{Generalized Koornwinder operators}
The theory is parameterized by six non-zero complex numbers $a,b,c,d,q,t$. Let
\begin{equation}\label{notations}
\sigma=\sqrt{\frac{a b c d}{q}},\quad t^\rho =(\sigma t^{N-i})_{i=1}^N .
\end{equation}
The following function takes the place of $\Delta$ in type $A$:
\begin{equation} \label{deltak}
\Delta^{(a,b,c,d)}(x)=\prod_{i=1}^N\prod_{\alpha\in\{a,b,c,d\}} \frac{(\frac{q}{x_i^2};q)_\infty}{(\frac{q}{\alpha x_i};q)_\infty
}
\prod_{1\leq i < j \leq N} \prod_{\epsilon=\pm1} \frac{(\frac{qx_j^\epsilon}{ x_i};q)_\infty}{(\frac{qx_j^\epsilon}{t  x_i};q)_\infty}.
\end{equation}
The Koornwinder difference operator\footnote{As in \cite{DFK21}, we add a constant to the original definition of \cite{Koornwinder}, which suits our choice of eigenvalues.} is defined as
\begin{equation}\label{korn}
\cD_1^{(a,b,c,d)}(\bx)= \sigma^2 \, {\rm Sym}\left\{ \Delta^{(a,b,c,d)}(\bx) \Gamma_1 \Delta^{(a,b,c,d)}(\bx)^{-1} \right\} +\varphi^{(a,b,c,d)}(\bx) 
\end{equation}
where the symmetrization is over the Weyl group $W$ and $\varphi^{(a,b,c,d)}(\bx)$ is a Weyl-symmetric rational function such that
$\cD_1^{(a,b,c,d)}(\bx)\cdot 1= \frac{1-t^N}{1-t}\left( 1+ \sigma^2 t^{N-1}\right)$. Explicitly, 
\begin{equation}\label{kormac}
\cD_1^{(a,b,c,d)}(\bx)=\frac{1-t^N}{1-t}\left( 1+ \sigma^2 t^{N-1}\right)+\sum_{i=1}^N\sum_{ \epsilon=\pm 1} \Phi_{i,\epsilon}^{(a,b,c,d)}(x) \, 
 \, (\Gamma_i^{\epsilon}-1)
\end{equation}
where \begin{eqnarray*}
\Phi_{i,\epsilon}^{(a,b,c,d)}(x) &=&\frac{(1-a x_i^{\epsilon})(1-b x_i^{\epsilon})(1-c x_i^{\epsilon})(1-d x_i^{\epsilon})}{(1-x_i^{2\epsilon})(1-q x_i^{2\epsilon})}\prod_{j\neq i} \frac{t x_i^\epsilon-x_j}{x_i^\epsilon-x_j}\frac{t x_i^\epsilon x_j-1}{x_i^\epsilon x_j-1}.\nonumber 
\end{eqnarray*}

The difference operator \eqref{kormac} is the first of a commuting family of $N$ difference operators, where the higher order operators can be chosen \cite{DFK21}, using \cite{vandiej,Rains}, so that the eigenvalues of their common eigenfunctions, the monic Koornwinder polynomials
$P_\lambda^{(a,b,c,d)}(\bx)$ indexed by integer partitions coding their leading term, are
\begin{equation}\label{eigenk} 
\cD_m^{(a,b,c,d)}(\bx)\,P_\lambda^{(a,b,c,d)}(\bx)=\sigma^m\,t^{m(N-\frac{m+1}{2})}\, {\hat e}_m(\bs)\, P_\lambda^{(a,b,c,d)}(\bx) ,
\end{equation}
for $m=1,2,...,N$, 
where $\bs=q^\lambda t^\rho$, with $\sigma,\rho$ as in \eqref{notations}.

Upon  specialization of the parameters $a,b,c,d$ (see Table \ref{table}), 
the Koornwinder theory reduces to the Macdonald theory \cite{macdoroot} for affine and twisted classical types. 

\begin{table}\label{table}
\begin{center}
\renewcommand{\arraystretch}{2}
\begin{tabular}{ |c|c|c|c|c|c|c|c|c|c|} 
\hline
 $\g$ & $\g^*$& $a$ & $b$ & $c$ & $d$ & $R$  & S & $R^*$ & $\xi_\g$ \\ 
\hline
$D_{N}^{(1)}$ & $D_{N}^{(1)}$ &$1$ & $-1$ & $q^{\frac{1}{2}}$ & $-q^{\frac{1}{2}}$ & $D_N$ & $D_N$ & $D_N$ & 0 \\ 
\hline
$B_{N}^{(1)}$ & $C_{N}^{(1)}$ &$t $ & $-1$ & $q^{\frac{1}{2}}$ & $-q^{\frac{1}{2}}$ & $B_N$ & $B_N$ & $C_N $ &$\frac12$\\ 
\hline
$C_{N}^{(1)}$ &$B_{N}^{(1)}$ &$t^{\frac{1}{2}}$ & $-t^{\frac{1}{2}}$ & $t^{\frac{1}{2}}\,q^{\frac{1}{2}}$ & $-t^{\frac{1}{2}}\,q^{\frac{1}{2}}$ & $C_N$ & $C_N$ &$B_N$& 1 \\
\hline
$A_{2N-1}^{(2)}$ & $A_{2N-1}^{(2)}$ &$t^{\frac{1}{2}}$ & $-t^{\frac{1}{2}}$ & $q^{\frac{1}{2}}$ & $-q^{\frac{1}{2}}$ & $C_N$ & $B_N$ &$C_N$& $\frac12$ \\
\hline
$D_{N+1}^{(2)}$ & $D_{N+1}^{(2)}$ &$t $ & $-1$ & $t\, q^{\frac{1}{2}}$ & $-q^{\frac{1}{2}}$ & $B_N$ & $C_N$ &$B_N$& 1\\
\hline
$A_{2N}^{(2)}$ & $A_{2N}^{(2)}$ &$t $ & $-1$ & $t^{\frac{1}{2}}\,q^{\frac{1}{2}}$ & $-t^{\frac{1}{2}}\,q^{\frac{1}{2}}$ & $BC_N$ & -- & $BC_N$& 1\\
\hline
\end{tabular}
\vskip.5cm
\caption{Specialization of the Koornwinder parameters $a,b,c,d$ corresponding to the affine algebra $\g$. The pair  $(R,S)$ refer to a pair of classical root systems corresponding to Macdonald's notation \cite{macdoroot}, except for $A_{2n}^{(2)}$. }\label{tableone}    \label{korspec}
\end{center}
\end{table}

\subsection{Duality and Pieri rules}
The Koornwinder polynomials obey a generalized version \cite{vDselfdual,Sahi} of the duality property for Macdonald polynomials \eqref{duapol}.
Let $*$ denote the following involution on the parameters $(a,b,c,d)$:
\begin{equation}
a^*=\Big(\frac{abcd}{q}\Big)^{1/2},\quad
b^*=-\Big(q\frac{ab}{cd}\Big)^{1/2},\quad
c^*=\Big(q\frac{ac}{bd}\Big)^{1/2},\quad
d^*=-\Big(q\frac{ad}{bc}\Big)^{1/2}.\label{dua}
\end{equation}
This implies that $\sigma^*=a$, and $(\sigma\,t^{\rho})^*=a\, t^\rho$.
The duality property for Koornwinder polynomials is 
\begin{equation}\label{duapolsabcd}
\frac{P_\lambda^{(a,b,c,d)}(q^\mu t^{\rho^*})}{P^{(a,b,c,d)}_\lambda(t^{\rho^*})}=
\frac{P_\mu^{(a^*,b^*,c^*,d^*)}(q^\lambda t^{\rho})}{P_\mu^{(a^*,b^*,c^*,d^*)}(t^{\rho})}.
\end{equation}
where $\mu,\lambda$ are both integer partitions. 
\begin{theorem}{\rm \cite{DFK21}} The denominators in \eqref{duapolsabcd} can be expressed
in terms of $\Delta^{(a,b,c,d)}(\bx)$ of \eqref{deltak} as:
\begin{eqnarray*}
P_\lambda^{(a,b,c,d)}(t^{\rho^*})&=&t^{(\rho^*,\lambda)}\frac{\Delta^{(a^*,b^*,c^*,d^*)}(t^{\rho})}{\Delta^{(a^*,b^*,c^*,d^*)}(q^\lambda t^{\rho})},\nonumber \\ 
P_\mu^{(a^*,b^*,c^*,d^*)}(t^{\rho})&=&t^{(\rho,\mu)}\frac{\Delta^{(a,b,c,d)}(t^{\rho^*})}{\Delta^{(a,b,c,d)}(q^\mu t^{\rho^*})} .
\end{eqnarray*}
\end{theorem}

As in Macdonald theory, Equation \eqref{duapolsabcd} can be generalized to the Universal solutions of the eigenvalue equations,
$P^{(a,b,c,d)}(\bx;\bs)=q^{(\lambda,\mu)}\sum_{\al\in Q_+} c_\al^{(a,b,c,d)}(\bs)\, \bx^{-\al} $ with generic $\bx=q^\mu t^{\rho^*}$ and $\bs= q^\lambda t^\rho$ with
$c_0^{(a,b,c,d)}(\bs)=1$.
The series $P^{(a,b,c,d)}(\bx;\bs)$ reduces to a Koornwinder polynomial upon specializing $\lambda$ to an integer partition\footnote{In the specialized cases of Table \ref{table}, further truncations occur when $\lambda$ are taken to be integral weights of the root system $R$, recovering the full generalized Macdonald polynomial content. }:
\begin{equation}\label{recoverk}
P^{(a,b,c,d)}(\bx;\bs)\Big\vert_{\bs=q^\lambda\, t^\rho,\, \lambda\, {\rm integer}}=t^{-(\rho,\lambda)}\,P_\lambda^{(a,b,c,d)}(\bx) 
\end{equation} 
\begin{theorem}{\rm \cite{DFK21}}
The universal solutions satisfy the duality relations
\begin{equation}\label{duaunivk}
\Delta^{(a^*,b^*,c^*,d^*)}(\bs)\, P^{(a,b,c,d)}(\bx;\bs)=\Delta^{(a,b,c,d)}(\bx)\, P^{(a^*,b^*,c^*,d^*)}(\bs;\bx) .
\end{equation}
\end{theorem}
Again, this implies the Pieri rules for Koornwinder polynomials: 
\begin{equation}
\label{pieriko}
\hat e_m(\bx)\, P_\lambda^{(a,b,c,d)}(\bx)=\cH_m^{(a,b,c,d)}(\bs) \, P_\lambda^{(a,b,c,d)}(\bx).
\end{equation}
The Pieri operator is computed from the generalized Koornwinder polynomials for the dual parameters:
\begin{equation}\label{opierik}
\cH_m^{(a,b,c,d)}(\bs)= a^{-m} t^{m(\frac{m+1}{2}-N)} \, {\rm Ad}_{\Delta^{(a^*,b^*,c^*,d^*)}(\bs)^{-1}\,t^{(\rho^*,\lambda)}} \left\{\cD_m^{(a,b,c,d)}(\bs)\right\}.
\end{equation}

\section{Duality for genus two generalized Macdonald theory}

The genus-2 DAHA of ``rank-1" was introduced recently by Arthamonov and Shakirov  using refined Chern-Simons theory \cite{ArtShak1,ArtShak}.
In the following, we provide a formulation of duality analogous to the 
discussion above.
Although this case is not directly related to a particular root lattice, note that the following vectors play an analogous role here.
Define the three vectors in $\C^3$:
$$\al_1=(1,1,-1),\quad \al_2=(1,-1,1), \quad \al_3=(-1,1,1) ,$$
and the root system-type sets $R_\pm^{(2)}$:
$$R_+^{(2)}=\{\al_1,\al_2,\al_3,\al_1+\al_2,\al_1+\al_3,\al_2+\al_3,\al_1+\al_2+\al_3\}=-R_-^{(2)}$$
together with the positive cone $Q_+^{(2)}=\oplus \Z_{\geq 0} \al_i$.
The (un-normalized) dual basis is denoted by
$$\omega_1=(1,1,0),\quad \omega_2=(1,0,1), \quad \omega_3=(0,1,1) .$$
The analog of the Weyl vector is taken as
$$\rho=\al_1+\al_2+\al_3=(1,1,1) .$$
The analog of integer partitions are the elements in the set
$$
\mathcal P^{(2)} = \left\{\lambda=(\lambda_1,\lambda_2,\lambda_3)\in \Z^3 | (\al_i,\lambda)\in 2 \Z_{\geq 0}, i=1,2,3\right\}.
$$
We use the notation $\bx=(x_1,x_2,x_3)$. The analog of the Weyl group (generated by reflections by $\alpha_i+\alpha_j$) is the product group $W\simeq S_2\times S_2\times S_2$, acting on the $\bx$-variables via inversions $x_i\mapsto x_i^{\pm1}$.

\subsection{Difference operators}
Let $q,t$ be two nonzero complex parameters.\footnote{We use different conventions from \cite{ArtShak} to avoid fractional powers of $q,t$. The results of \cite{ArtShak} are recovered upon substituting $q\to q^{1/2}$ and $t\to t^{1/2}$.}
Introduce the infinite product:
\begin{eqnarray*}
\Delta^{(2)}(\bx)&=&\frac{\displaystyle\prod_{1\leq i<j\leq 3}(q^2\bx^{-\al_i-\al_j} ;q^2)_\infty} {\displaystyle({q^2}{t^{-1}} \bx^{-\al_1-\al_2-\al_3};q^2)_\infty \prod_{i=1}^3 ({q^2}{t^{-1}} \bx^{-\al_i};q^2)_\infty}\\
&=&
\prod_{n=1}^{\infty} \frac{1}{\displaystyle1-\frac{q^{2n}}{t \, x_1x_2x_3}} \prod_{i=1}^3 \frac{\displaystyle1-\frac{q^{2n}}{x_i^2} }{\displaystyle1-\frac{q^{2n}}{t} \frac{x_i^2}{x_1 x_2 x_3}}.
\end{eqnarray*}

The genus-2  commuting operators are defined for each pair $(i,j)$ such that  $1\leq i<j\leq 3$  by\footnote{This definition leads to the {\it same} difference operators as those introduced in \cite{ArtShak}.}
\begin{eqnarray}
\label{mactwo}
\cD_{i,j}^{(2)}&=&t\, {\rm Sym} \left\{ \Delta^{(2)}(\bx)\, \Gamma_i\Gamma_j \, \Delta^{(2)}(\bx)^{-1} \right\} \nonumber \\ &=&\frac{1}{t}\sum_{\epsilon_i,\epsilon_j=\pm 1} 
\frac{(t x_i^{\epsilon_i}x_j^{\epsilon_j}x_k-1)(t x_i^{\epsilon_i}x_j^{\epsilon_j}x_k^{-1}-1)}{(x_i^{2\epsilon_i}-1)(x_j^{2\epsilon_j}-1)} \Gamma_i^{\epsilon_i}\Gamma_j^{\epsilon_j},
\end{eqnarray}
where $k$ is such that $\{i,j,k\}=\{1,2,3\}$. Here, the symmetrization is with respect to $W$.

The genus 2 Macdonald polynomials $P_\lambda^{(2)}(\bx)$ are the monic\footnote{A different normalization was used in the original definition \cite{ArtShak} where 
the polynomials are not monic, but inherit a particular normalization from the refined Chern-Simons perspective.}
common eigenfunctions of the three operators $\cD_{i,j}^{(2)}$, indexed by genus-2 partitions $\lambda\in \mathcal P^{(2)}$ coding their leading term in the sense that 
$$P_\lambda^{(2)}(\bx)=\prod_{i=1}^3 x_i^{\frac12 (\al_i,\lambda)} \Big(1+O(\bx^{-\al})_{\al\in Q_+}\Big). $$

The corresponding eigenvalue equations  for $1\leq i<j\leq 3$ are
\begin{equation}\label{eval2}
\cD_{i,j}^{(2)}(\bx) \, P_\lambda^{(2)}(\bx)= \hat e_1(s_{\ell})\, P_\lambda^{(2)}(\bx)
\end{equation} 
where $\ell$ is such that $\al_i+\al_j=2\epsilon_\ell$, $s_\ell=t\, q^{\frac12 (\al_i+\al_j,\lambda)}=t\,q^{\lambda_\ell}$, and $\hat e_1(x)=x+x^{-1}$. 
Note that we can still write $\bs=q^{\lambda}t^\rho$, with the relevant definitions above.

\subsection{Duality and Pieri rules}

In  \cite{ArtShak}, the Macdonald polynomials of genus-2 are introduced as the unique solution to Pieri rules, which automatically fixes their normalization. These were inferred from the refined Chern-Simons approach. We now show that the Pieri rules for the monic normalization of genus-2 Macdonald polynomials can be derived as a consequence of duality relations. The resulting Pieri rules agree with those of  \cite{ArtShak} up to a suitable change of normalization. We do not present the proof here for lack of space, see
\cite{DFK23}. The the method of the proof is similar to that used by Macdonald for type $A$ \cite{macdo}.

\begin{theorem} The genus 2 Macdonald polynomials obey the following duality property, obtained by specializing both $\bx= q^\mu t^\rho$ and $\bs=q^{\lambda}t^\rho $ with $\lambda,\mu\in\mathcal P^{(2)}$:
\begin{equation}\label{duapol2}
\frac{P_\lambda^{(2)}(q^{\mu} t^\rho)}{P_\lambda^{(2)}(t^\rho)}=\frac{P_\mu^{(2)}(q^{\lambda} t^\rho)}{P_\mu^{(2)}(t^\rho)} .
\end{equation}
\end{theorem}

\begin{theorem} The normalization factors for genus 2 Macdonald polynomials are expressed in terms of the product $\Delta^{(2)}(\bx)$ as follows:
\begin{eqnarray}
P_\lambda^{(2)}(t^\rho)=t^{\frac12 (\rho,\lambda)} \frac{\Delta^{(2)}(t^\rho)}{\Delta^{(2)}(q^{\lambda} t^\rho)} .
\end{eqnarray}
\end{theorem} 
As in genus-1, the duality extends to the universal solution of the eigenvalue 
equations \eqref{eval2}, defined for $\bx=q^\mu t^\rho,\bs=q^\lambda t^\rho$ with generic comples $\lambda$, $\mu$ as the series
$$P^{(2)}(\bx;\bs)= q^{\frac12 (\lambda,\Omega\mu)}\,\sum_{\al\in Q^{(2)}_+} c_\al^{(2)}(\bs)\, \bx^{-\al} , \qquad  \Omega=\begin{pmatrix}1 & 1 & -1\\
1 & -1 & 1\\
-1 & 1 & 1\end{pmatrix}, $$ 
with the normalization
$c^{(2)}_0(\bs)=1$. As before, the series has a finite truncation when we specialize $\lambda$ to a genus 2 integer partition, recovering:
\begin{equation}\label{recover2}
P^{(2)}(\bx;\bs)\Big\vert_{\lambda\in\mathcal P^{(2)}}=t^{-\frac12 (\rho,\lambda)}\,P_\lambda^{(2)}(\bx) ,
\end{equation}
where we used $\prod_{i=1}^3 x_i^{\frac12 (\al_i,\lambda)} =t^{\frac12 (\rho,\lambda)} q^{\frac12 (\lambda,\Omega\mu)}$. 
\begin{theorem}The duality relation \eqref{duapol2} extends to the universal solution with generic $\lambda,\mu$ as
\begin{equation}\label{duau2}
\Delta^{(2)}(\bs)\, P^{(2)}(\bx;\bs)= \Delta^{(2)}(\bx)\, P^{(2)}(\bs;\bx) .
\end{equation}
\end{theorem}

As in genus-1, duality implies Pieri rules for genus-2 Macdonald polynomials. The Pieri operators express the result of multiplying the polynomials by any of 
the three elementary symmetric functions:
\begin{equation}\label{pieri2}
\hat e_1(x_\ell)\, P_\lambda^{(2)}(\bx)=\cH_\ell^{(2)}(\bs) \, P_\lambda^{(2)}(\bx) ,\qquad \ell=1,2,3.
\end{equation}
Duality implies that the Pieri operators $\cH_\ell^{(2)}(\bs)$ are
\begin{equation}\label{pieriop2}
\cH_\ell^{(2)}(\bs)={\rm Ad}_{\Delta^{(2)}(\bs)^{-1}\,t^{\frac12 (\rho,\lambda)}} \left\{\cD_{i,j}^{(2)}(\bs)\right\},\qquad i<j,\ \alpha_i+\alpha_j=2\epsilon_\ell.
\end{equation}
Explicitly, using the notation $\Lambda_i=q^{\lambda_i}$ and $T_i=q^{\lambda_i\partial_{\lambda_i}}$, 
\begin{eqnarray*}
\cH_1^{(2)}(\bs)&=&T_1T_2+t \frac{\Lambda_1\Lambda_2}{\Lambda_3} 
\frac{(1-\frac{\Lambda_2\Lambda_3}{\Lambda_1})(1-q^{-2}t^{2}\frac{\Lambda_2\Lambda_3}{\Lambda_1})}{(1-t^2\Lambda_2^2)(1-q^{-2}t^2\Lambda_2^2)}
\, T_1T_2^{-1}\\
&&
+ t \frac{\Lambda_1\Lambda_2}{\Lambda_3} 
\frac{(1-\frac{\Lambda_1\Lambda_3}{\Lambda_2})(1-q^{-2}t^{2}\frac{\Lambda_1\Lambda_3}{\Lambda_2})}{(1-t^2\Lambda_1^2)(1-q^{-2}t^2\Lambda_1^2)}
\, T_1^{-1}T_2\\
&&
+\frac{(1-\frac{\Lambda_1\Lambda_2}{\Lambda_3})(1-q^{-2}t^{2}\frac{\Lambda_1\Lambda_2}{\Lambda_3})(1-t^2\Lambda_1\Lambda_2\Lambda_3)(1-q^{-2}t^4\Lambda_1\Lambda_2\Lambda_3)}{(1-t^2\Lambda_1^2)(1-t^2\Lambda_2^2)(1-q^{-2}t^2\Lambda_1^2)(1-q^{-2}t^2\Lambda_2^2)} T_1^{-1}T_2^{-1},
\end{eqnarray*}
and $\cH_2^{(2)}(\bs)$ and $\cH_3^{(2)}(\bs)$ are obtained by cyclic permutations of the indices $1,2,3$.

These Pieri rules agree with those of \cite{ArtShak} for the polynomials $\psi_\lambda(\bx)$  with the normalization
$$ \psi_\lambda(\bx)=t^{-(\rho,\lambda)}\,\prod_{k=1}^3 \prod_{i=0}^{\lambda_k-1} \frac{1-t^4 q^{2i}}{1-t^2 q^{2i}} \, \frac{P_\lambda(\bx)}{P_\lambda(t^\rho)} .$$

\section{$q$-Whittaker limits and the quantum Q-systems}
In genus-1, the motivation for the construction of the particular sets of difference operators referred to above above and their Dehn-twisted analogues is a set of recursion relations for characters of finite-dimensional representations of quantum affine algebras known as the Q-system \cite{HKOTY} and its quantization \cite{DFKqKR}.  These appear in the $q$-Whittaker limit of the sDAHAs.
\subsection{Genus-1}
The functions $\Delta$ for each setting above were chosen so that the limit $t\to\infty$ is well-defined. We refer to this as the $q$-Whittaker limit. Each of the $q$-difference operators in $\bx$ has a well-defined limit (after renormalization by a power of $t$). 
This limit can be taken directly in Macdonald theory for type $A_N^{(1)}$. In Koornwinder theory, we take the limit after the specialization of the parameters $(a,b,c,d)$ as in Table Table \ref{table}, since the specialization depends on $t$.

In this limit, the Dehn-twisted Macdonald-Koornwinder difference operators do not simplify greatly, for example \eqref{translatedmacdonald} become
$$ D_{a,n}(\bx)=\lim_{t\to\infty} t^{-a(N-a)} \cD_{a,n}(\bx)=\sum_{I\subset [1,N]\atop |I|=a} 
\prod_{i\in I,j\not\in I} \frac{x_i}{x_i-x_j} \, \prod_{i\in I} x_i^n \, \prod_{i\in I}\Gamma_i .$$

 The eigenvalues of the Macdonald-Koornwinder operators, which are elementary symmetric functions in $\bs$, are replaced by their dominant monomial in the limit, because $\bs$ depends explicitly on $t$. For example in type $A$, $e_a(\bs)$ is replaced by $\Lambda^{\omega_a}$, where $\Lambda_i=q^{\lambda_i}$.
The corresponding eigenfunctions are (class-1) $q$-Whittaker functions, $\Pi_\lambda(\bx)$. In type $A$ they satisfy the equations
\begin{equation}\label{evalueqwhit}
 D_{a,0}(\bx)\, \Pi_\lambda(\bx) = \Lambda_1\Lambda_2\cdots \Lambda_a\, \Pi_\lambda(\bx) .
 \end{equation}
 
The universal solutions of the eigenvalue equations also have well-defined $q$-Whittaker limits as series solutions in $\bx^{-\alpha_i}$, $\Pi(\bx;\Lambda)$.  Because $\bs$ depends explicitly on $t$, the the symmetry  $\bx\leftrightarrow \bs$ in Equations \eqref{duauniv}, \eqref{duaunivk} and \eqref{duau2} is broken in the limit, $\Delta(\bs)\to 1$ whereas $\Delta(\bx)$ is an still an infinite product. 

The Pieri rules have well-defined
limits of the form $e_a(\bx)\,\Pi_\lambda(x)=H_a(\Lambda) \, \Pi_\lambda(\bx)$ in type $A$. 
The limits of the Pieri operators are greatly simplified in the limit and can be identified with the quantum relativistic Toda 
Hamiltonians associated with each root system. They are all Laurent polynomials in $\Lambda_i, T_i$. For example in type A they are
$$H_1(\Lambda)=T_1+\sum_{i=2}^N \Big(1-\frac{\Lambda_{i}}{\Lambda_{i-1}}\Big) T_i .$$
These Hamiltonians are the conserved quantities of the discrete evolutions called the quantum Q-systems corresponding to each affine algebra
\cite{DFK15,DFK21}. 
These are non-commutative rational recursion relations in the discrete time variables $n$. The limiting Dehn-twisted Macdonald-Koornwinder operators are the solutions to these equations. In most cases, the Dehn twist is a sequence of mutations in a quantum cluster algebra in which the elements $D_{a,n}$ are the $\mathcal A$-type cluster variables. The proof of these statements, found in \cite{DFK21}, relies on the duality property and the notion of the ``Fourier transform", which allows us to trade $q$-difference operators with their eigenvalues, when acting on the universal $q$-Whittaker function.

\subsection{Genus-2 case}

We use the same $t\to\infty$ limit for in genus-2. In this limit, the operators \eqref{mactwo} factorize into genus-1 $A_1^{(1)}$ type Macdonald operators:
\begin{equation}\label{split}
D_{i,j}^{(2)}(\bx)=\lim_{t\to\infty} t^{-1} \cD_{i,j}^{(2)}(\bx)
=\sum_{\epsilon_i,\epsilon_j=\pm 1} 
\frac{x_i^{2\epsilon_i}}{x_i^{2\epsilon_i}-1} \,\Gamma_i^{\epsilon_i}\, \frac{x_j^{2\epsilon_j}}{x_j^{2\epsilon_j}-1} \Gamma_j^{\epsilon_j}=D_1(x_i)\,D_1(x_j)
\end{equation}
$D_1(x)$ is the ${\mathfrak sl}_2$ Macdonald operator in the $q$-Whittaker limit
$$D_1(x)=\frac{x^{2}}{x_i^{2}-1}\Gamma_x +\frac{x^{-2}}{x_i^{-2}-1}\Gamma_x^{-1}=D_{1}(x,x^{-1}).$$

As a consequence, the  $q$-Whittaker limit of the genus-2 Macdonald polynomials factorize:
$$\Pi^{(2)}_\lambda(\bx)=\Pi_{\frac12 (\al_1,\lambda)}(x_1) \, \Pi_{\frac12 (\al_2,\lambda)}(x_2)\,\Pi_{\frac12 (\al_3,\lambda)}(x_3) ,$$
where $\Pi_m(x)$ are expressed in terms of the $A_1$ Macdonald polynomials as $\Pi_{\lambda_1-\lambda_2}(x)=\Pi_{\lambda_1,\lambda_2}(x,x^{-1})$.

The Pieri operators have simple limits as well:
$$H_1^{(2)}(\Lambda)=T_1T_2+\Big(1-\frac{\Lambda_3}{\Lambda_1\Lambda_2}\Big)T_1^{-1}T_2^{-1}, $$
while $H_2^{(2)},H_3^{(2)}$ are obtained by cyclic permutation of the indices. The latter coincides with the Toda Hamiltonian $H_1(\Lambda_1,\Lambda_2)$ 
for $A_1$ under the correspondence 
$$\Lambda_1\to \Big(\frac{\Lambda_1\Lambda_2}{\Lambda_3}\Big)^{1/2}=q^{\frac12 (\al_1,\lambda)},\quad 
\Lambda_2\to \Big(\frac{\Lambda_1\Lambda_2}{\Lambda_3}\Big)^{-1/2}=q^{-\frac12 (\al_1,\lambda)} .$$
The quantum $Q$-system cluster algebra in this case is a factorized system of three type $A_1^{(1)}$ systems, corresponding to three disconnected Jordan quivers.

\section{Conclusion}
We have summarized the duality property in several versions of the spherical double affine Hecke algebras. In each case the function $\Delta(\bx)$, which is related to the normalization of the Macdonald polynomials under duality, plays a pivotal role in determining the Pieri rules. This function is related to the affine root system of an underlying affine Lie algebra in the genus-1 case. We have argued that there is a structural similarity in the genus-2 case, which does not correspond to a root system. The proofs of the statements for genus-2 will appear in a forthcoming publication \cite{DFK23}.

 \bibliographystyle{alpha}
 \bibliography{refs}

\end{document}